\documentstyle[preprint,epsfig,aps]{revtex}
\begin{document}
\draft
\title{Phenomenological study of hadron interaction models} 

\author{H.R. Pang$^1$, J.L. Ping$^2$, Fan Wang$^1$ and T.Goldman$^3$}
\address{$^1$Physics Department, Nanjing University, Nanjing, 210093, China}
\address{$^2$Physics Department, Nanjing normal University, Nanjing, 210097, 
China}
\address{$^3$Theoretical Division, LANL, Los Alamos, NM 87545, USA} 
\maketitle
\begin{abstract}

We present a phenomenological study of three models with different
effective degrees of freedom: a Goldstone Boson Exchange (GBE) model
which is based on quark-meson couplings, the quark delocalization,
color screening model (QDCSM) which is based on quark-gluon couplings
with delocalized quark wavefunctions, and the Fujiwara-Nijmegen (FN)
mixed model which includes both quark-meson and quark-gluon couplings.
We find that for roughly two-thirds of 64 states consisting of pairs of
octet and decuplet baryons, the three models predict similar effective
baryon-baryon interactions. This suggests that the three very different
models, based on different effective degrees of freedom, are
nonetheless all compatible with respect to baryon spectra and
baryon-baryon interactions. We also discuss the differences between the
three models and their separate characteristics.

\end{abstract}


\begin{flushright}
\vspace*{-8.2in}
LA-UR-01-1908
\end{flushright}
\setcounter{page}{0}

\pagebreak

\section{introduction}
Hadronic interactions are central to strong interaction physics.
However,  for the time being, the fundamental strong interaction
theory, Quantum ChromoDynamics (QCD) remains too complicated for these
to be directly calculated from it.  Nevertheless, initial lattice QCD
calculations of hadronic interactions\cite{dr} have been initiated. Due
to the complication of a multiquark system, however, quantitative
results remain in the distant future.

Meson exchange models, based on meson-baryon couplings, were developed
long before QCD\cite{yukawa} and are still the best at fitting the vast
collection of NN experimental data\cite{mach}.  Unfortunately, the QCD
basis for such effective degrees of freedom in {\em t-channel}
exchanges is not clear at present. Chiral perturbation effective field
theory\cite{wein} is well-based in QCD; for a recent derivation, see
\cite{wangq}. However, due to the fact that the quark and gluon degrees
of freedom have been integrated out, it can not be used to study the
quark-gluon internal structure of hadrons. Nor is it suitable for the
study of genuine multiquark-gluon systems such as glueballs, hybrids,
$q\bar{q}q\bar{q}$ systems, and dibaryons.

R.T.Cahill et al.\cite{cah} have developed an effective field theory
which takes spontaneous chiral symmetry breaking into account.
Constituent quarks and Goldstone bosons appear here as the effective
degrees of freedom for low energy QCD physics. This model has been
applied both to pion and $\sigma$ meson internal structure and to meson
interactions, but not yet to NN interactions\cite{caht}. L.Ya.Glozman,
D.O.Riska and G.E.Brown\cite{grb} propose a phenomenological model,
with constituent quarks and Goldstone bosons as the effective degrees
of freedom for describing baryon spectroscopy and baryon interactions.
However, their quark meson coupling is the linear Yukawa coupling,
which reflects neither QCD based chiral perturbation nor the nonlinear
coupling obtained from the effective field theory referred to above.

A.Manohar and H.Georgi\cite{mg} have argued that below the chiral
symmetry breaking scale($\sim$1Gev) and within the confinement
regime ($\sim$1fm), the proper effective degrees of freedom are
Goldstone bosons, constituent quarks and gluons.(We have constructed
an approximate QCD derivation of this model, similar to Cahill's 
approach, but the quantum fluctuations of the gluon field are suppressed.)
Such a hybrid quark-gluon-meson exchange model has been widely used to
describe nucleon-baryon interactions and a semi-quantitative fit has
been obtained\cite{tff}.

Models with constituent quarks and effective one gluon
exchange\cite{rggi} describe hadron spectroscopy quite well. The
Hamiltonian approach to QCD\cite{gji} seeks the QCD basis for this
model.  The Schwinger-Dyson equation approach also keeps constituent
quarks and gluons as the effective degrees of freedom\cite{sch}.
However, direct extension of this model to the NN interaction obtains
only a repulsive core\cite{cww}.  A variant of this model obtains the
NN repulsive core and intermediate range attraction simultaneously, by
taking into account both quark delocalization and color screening. This
also describes a qualitative similarity between nuclear and molecular
forces. A semiquantitative fit to the existing NN, N$\Lambda$, and
N$\Sigma$ scattering data and to properties of the deuteron has been
obtained as well \cite{wpwg}.

The MIT bag model uses current quarks and gluons to describe hadron
internal structure\cite{jaffe}. A quark-meson coupling has to be
introduced to restore chiral symmetry\cite{thomas} to this model;
Cahill et al. have given an approximate QCD derivation\cite{cah}.  The
model has been extended to the description of hadron interactions using
the R-matrix method and compound quark model approach\cite{lomon}.

The Skyrmion model can be viewed as originating from the large $N_c$
limit. It has been used both for nucleon internal structure and for
hadronic interactions\cite {izah}. The nontopological soliton model
also acquired an approximate QCD derivation a few years
ago\cite{twen}.

There might be even more QCD models that can be listed. But this short
review suggests that neither theoretical studies of QCD nor
phenomenological model analyses can yet determine which effective degrees
of freedom are best, nor which model is best for the description of low
energy hadron physics. Therefore, theoretical study of QCD
and phenomenological analysis are both still needed to extend our
understanding of low energy strong interaction physics.

This paper reports a phenomenological study of the effective
baryon-baryon interactions of the ground state octet and decuplet
baryons using three constituent quark models with quark-meson,
quark-meson-gluon, and quark-gluon effective degrees of freedom
respectively. The results show that in many channels, (about 2/3 of the
total states) the three different models give more or less the same
predictions.  We take this as a phenomenological verification that,
even for baryon-baryon interactions, quark models with different
effective degrees of freedom are quite compatible and that meson
exchange effects are modeled, at least to some extent, by quark
delocalization and color screening. In some channels, however,
especially promising dibaryon states, the different models give
drastically different predictions. These cannot be viewed as being as
reliable as those above; i.e., these results are sensitive to model
details. Experimental searches in those channels are needed to
distinguish the validity of the different models.

This paper is organized as follows: In Sec. II, we describe the
Hamiltonians and parameters of the three models.  In Sec. III, we
present the results and discuss them.  The final section gives our
conclusions.

\section{Hamiltonian and parameters of three models}

From among the models discussed in the introduction, we choose three
constituent quark models on which to do a comparative phenomenological
study; i.e., we calculate the adiabatic effective baryon-baryon
interactions and dibaryon candidates systematically. The three models
are:  the Glozman-Riska-Brown model, based on constituent quarks and
exchange of Goldstone bosons (GBE); the Fujiwara model, based on
constituent quarks and gluons, with the Nijmegen description of one
boson exchange (FN); the quark delocalization and color screening model
(QDCSM), based on constituent quarks and gluons but with quark
delocalization and color screening effects included.  We choose to
consider these three models for several reasons: There has been
considerable debate recently regarding these effective degrees of
freedom\cite{twenone}; These models have been extensively studied with
respect to baryon spectroscopy and baryon-baryon interactions; They can
be calculated straightforwardly in a consistent, systematic way, using
fractional parentage expansions and variational methods.

\subsection{Quark and meson effective degrees of freedom: GBE models}

We take the Goldstone Boson Exchange (GBE) model of Glozman's group as
the example with constituent quarks and Goldstone bosons as the
effective degrees of freedom. This model has been successfully applied
to baryon spectra and also extended to study NN interactions. Although
recent research shows that scalar and vector mesons should also be
included to describe baryon-baryon interactions, here we use only the
original version, i.e., only pseudo-scalar mesons (as Goldstone
bosons), which gives a surprisingly good description of baryonic
spectra.

In the nonrelativistic meson-exchange quark-quark interaction of any
such model, there is a $\delta$-function term, which is true only for
point-like particles. For particles with structure, this
$\delta$-function must be smeared out. In the GBE model, two types of
smearing were used: Gaussian and Yukawa. We designate them as model
GBE(a) and GBE(b), respectively.

\subsubsection{GBE with Gaussian smearing: GBE(a)}

In this case, the $\delta$-function is smeared by a Gaussian
function\cite{light}. This version provides a successful description of
the spectroscopy of baryons. In addition, the short-range repulsion of
the NN interaction is also obtained.  This Gaussian smearing involves
two parameters corresponding to the short range cutoff point ($r_0$)
and the width ($\alpha$) of a bell-shaped curve.

\begin{eqnarray}
H&=&\sum_im_i+\sum_i\frac{P_i^2}{2m_i}-\frac{{(\sum_iP_i)}^2}{2\sum_im_i}
+\sum_{i<j}V_{conf}(r_{ij})+\sum_{i<j}V_{\chi}(r_{ij}) \nonumber \\
V_{conf}(r_{ij})&=&-\frac{3}{8}{\lambda}_i^c\cdot{\lambda}_j^c
(Cr_{ij}+V_0)\nonumber \\
V_{\chi}(r_{ij})&=&\{\sum_{F=1}^3V_{\pi}(r_{ij})
{\lambda}_i^F\cdot{\lambda}_j^F+
\sum_{F=4}^7 V_k(r_{ij}){\lambda}_i^F\cdot{\lambda}_j^F+
V_{\eta}(r_{ij}){\lambda}_i^8\cdot{\lambda}_j^8+V_{{\eta}^{'}}(r_{ij})
{\lambda}_i^0\cdot{\lambda}_j^0\}{\sigma}_i\cdot{\sigma}_j \nonumber \\
V_{\gamma}(r)&=&\frac{g^2_{\gamma}}{4\pi}\frac{1}{12m_im_j}\{
{\theta}(r-r_0){\mu}^2_{\gamma}\frac{e^{-{\mu}_{\gamma}r}}{r}
-\frac{4}{\sqrt{\pi}}{\alpha}^3exp(-{\alpha}^2{(r-r_0)}^2)\} \nonumber \\
&&{\gamma}=\pi,k,{\eta},{\eta}^{'} \nonumber 
\end{eqnarray}
where the parameters are

\[{\mu}_{\pi}=139 MeV, {\mu}_k=495 MeV, {\mu}_{\eta}=547 MeV,
 {\mu}_{{\eta}^{'}}=958 MeV\]
\[m_{u,d}=340 MeV, m_s=440 MeV,
  \frac{g^2_{{\pi}q}}{4\pi}=\frac{g^2_{{\eta}q}}{4\pi}
  =\frac{g^2_{kq}}{4\pi}=0.67\]
\[\frac{g^2_{{\eta}^{'}q}}{4\pi}=1.206, r_0=0.43 fm, 
 \alpha=2.91{fm}^{-1},V_0=0,C=0.474{fm}^{-2},b=0.437fm. \]
Here $b$ is the baryon size parameter (a parameter of the quark 
orbital Gaussian wave function).

\subsubsection{GBE with Yukawa smearing: GBE(b)}

Because Gaussian smearing does not meet the well known requirement that
the volume integral of the pseudoscalar meson exchange interaction
should vanish, another smearing version, Yukawa smearing, was also
applied in the GBE model\cite{nucleon}.

\[V_{\gamma}(r)=\frac{g_{\gamma}^2}{4\pi}\frac{1}{12m_im_j}
({\mu}_{\gamma}^2\frac{e^{-{\mu}_{\gamma}r}}{r}
-{\Lambda}_{\gamma}^2\frac{e^{-{\Lambda}_{\gamma}r}}{r}),
{\Lambda}_{\gamma}={\Lambda}_0+{\kappa}{\mu}_{\gamma}\]
\[m_{u,d}=340MeV, m_s=507MeV, C=0.77{fm}^{-2},\]
\[\frac{g_{{\pi}q}^2}{4\pi}=
\frac{g_{{K}q}^2}{4\pi}=
\frac{g_{{\eta}q}^2}{4\pi}=1.24~~,
~~\frac{g_{{\eta'}q}^2}{4\pi}=2.7652,
{\Lambda}_0=5.82{fm}^{-1},\kappa=1.34 \]
\[b=0.537fm, V_0=-686.4MeV\]

The values of $b$ and $V_0$ are obtained by reproducing the $N-\Delta$
mass difference and the nucleon mass with a Gaussian single quark
orbital wave function, as is used in standard quark model
calculations. We also assume a single quark-meson coupling constant,
$\frac{g_8^2}{4\pi}$, for all octet mesons ($\pi$, $K$, $\eta$) as the
originators of the model have done in their calculations.  The strange
quark mass, $m_s$, is determined by an overall fit to the masses of the
strange baryons.  Since "irrespective of the parametrization, the
flavor-spin symmetry is essential in the model"\cite{nucleon}, we
presume that we have maintained the essence of the GBE.  In addition,
only the spin-spin component of the pseudoscalar exchange interaction
has been included in the calculation.  We note that there are
extensions of the GBE which include pseudoscalar, vector and scalar
meson exchanges with all possible force components\cite{extension}.
This is expected to be necessary for baryon-baryon interactions, but
this addition will be left for future studies.

\subsection{Quark-meson-gluon effective degrees of freedom: FN model}

The naive quark-gluon exchange model is quite successful in baryon
spectroscopy, but its extension to baryon-baryon interactions is much
less so.  Only a repulsive core is obtained, but no intermediate range
attraction develops in the best studied NN channel.  Meson exchange
interactions are introduced into the model to provide this well
established attraction and the resulting hybrid model with quark, meson
and gluon effective degrees of freedom is widely used in baryon
interaction calculations.  There are several different model variations
in this approach.  We choose Fujiwara's model (FN)\cite{des} as an
example since it gives a simultaneous description of both NN and NY
(nucleon-hyperon) interactions.

\begin{eqnarray}
H&=&{\sum}^{6}_{i=1}(m_i+\frac{P_i^2}{2m_i})
+{\sum}^6_{i<j}(U_{ij}^{Cf}+U_{ij}^{FB}+{\sum}_{\beta}U_{ij}^{S\beta}
+{\sum}_{\beta}U_{ij}^{PS\beta}) \nonumber \\
U_{ij}^{Cf}&=&-a_c{\lambda}_i^C\cdot{\lambda}_j^Cr_{ij}^2 \nonumber \\
U_{ij}^{FB}&=&\frac{{\alpha}_s}{4}{\lambda}_i^C\cdot{\lambda}_j^C
\{\frac{1}{r_{ij}}
-\frac{\pi}{2}(\frac{1}{m_i^2}+\frac{1}{m_j^2}+\frac{4}{3}
\frac{{\sigma}_i\cdot{\sigma}_j}{m_im_j})\delta(\vec{r_{ij}})
-\frac{1}{2m_im_j}(\frac{\vec{P_i}\cdot\vec{P_j}}{r_{ij}}
+\frac{\vec{r_{ij}}\cdot(\vec{r_{ij}}\cdot\vec{P_i})\vec{P_j}}{r_{ij}^3})\}
\nonumber \\
U_{ij}^{PS\beta}&=&w_{ij}^{PS\beta}{(\frac{m_{\beta}}{m_{{\pi}^{+}}})}^2
\frac{m_{\beta}}{3}
\{{\sigma}_i\cdot{\sigma}_j[Y(x)-c_{\delta}\frac{4\pi}{m_{\beta}^3}
\delta(\vec{r_{ij}})]\} \nonumber \\
U_{ij}^{S\beta}&=&-w_{ij}^{S\beta}m_{\beta}Y(x) \nonumber \\
&&Y(x)=\frac{e^{-x}}{x},~~~x=m_{\beta}|r_{ij}| \nonumber \\
&&w_{ij}^{\pi}={(\frac{3}{5}f_8^{PS})}^2{\sum}_{F=1}^3{\lambda}_i^F
{\lambda}_j^F
,~~~w_{ij}^K={(\frac{3}{5}f_8^{PS})}^2{\sum}_{F=4}^7{\lambda}_i^F
{\lambda}_j^F
\nonumber \\
&&w_{ij}^{\eta}=(-f_1^{PS}sin{\theta}_{PS}+\frac{3}{5}f_8^{PS}
cos{\theta}_{PS}
{\lambda}_i^8)\times(-f_1^{PS}sin{\theta}_{PS}+\frac{3}{5}f_8^{PS}
cos{\theta}
_{PS}{\lambda}_j^8), \nonumber \\
&&w_{ij}^{{\eta}^{'}}=(f_1^{PS}cos{\theta}_{PS}+\frac{3}{5}f_8^{PS}
sin{\theta}_{PS}
{\lambda}_i^8)\times(f_1^{PS}cos{\theta}_{PS}+\frac{3}{5}f_8^{PS}
sin{\theta}_{PS}{\lambda}_j^8) \nonumber \\
&&w_{ij}^{\delta}={(f_8^S)}^2{\sum}_{F=1}^3{\lambda}_i^F{\lambda}_j^F,~~~
w_{ij}^{\kappa}={(f_8^S)}^2{\sum}_{F=4}^7{\lambda}_i^F{\lambda}_j^F,
\nonumber \\
&&w_{ij}^{\epsilon}=(\frac{1}{3}f_1^Scos{\theta}_S+f_8^Ssin{\theta}_S
{\lambda}_i^8
)\times(\frac{1}{3}f_1^Scos{\theta}_S+f_8^Ssin{\theta}_S{\lambda}_j^8),
\nonumber \\
&&w_{ij}^{S*}=(-\frac{1}{3}f_1^Ssin{\theta}_S+f_8^Scos{\theta}_S
{\lambda}_i^8
)\times(-\frac{1}{3}f_1^Ssin{\theta}_S+f_8^Scos{\theta}_S{\lambda}_j^8)
\nonumber 
\end{eqnarray}
\[m_{{\pi}^{+}}=140MeV, m_{\pi}=138MeV, m_K=496MeV, m_{\eta}=547MeV, m_{{\eta}^{'}}=958MeV,\]
\[m_{\delta}=970MeV, m_{\kappa}=1145MeV, m_{\epsilon}=800MeV, m_{S*}=1250MeV,
c_{\delta}=0.381\]
\[f_1^S=2.89138, f_8^S=1.07509, f_1^{PS}=0.21426, f_8^{PS}=0.26994, {\theta}_{PS}=-23^{0},{\theta}_S=27.78^{0}\]
\[b=0.616fm, a_c=95.61MeV{\cdot}fm, m_{ud}=360MeV, {\alpha}_s=2.1742, {\lambda}=m_s/m_{ud}=1.526\]

\subsection{Quark-gluon effective degrees of freedom: QDCSM}

For this case, we take the quark delocalization, color screening model
(QDCSM) as the example\cite{wpwg}. As noted above, the naive quark
model with quark and gluon effective degrees of freedom gives a good
description of individual baryon properties. However, its extension to
baryon-baryon interactions is not successful.  The QDCSM maintains
quarks and gluons as the effective degrees of freedom, but enlarges the
Hilbert space by incorporating quark delocalization. It also
distinguishes the confinement potential between quark pairs inside one
baryon and from different baryons by introducing color screening. This
is based on the recognition that the interaction between quarks from
different color singlet baryons must be screened as color neutrality is
observed to be maintained on a $fm$ scale. To be precise, the backflow
of color necessary is not explicitly modelled, but is accomplished by an
effective matrix elements method.  The QDCSM gives a reasonable fit to
NN, N$\Lambda$ and N$\Sigma$ scattering data.

\begin{eqnarray} 
H(6) &=& \sum_{i=1}^{6}(m_{i}+\frac{p_{i}^{2}}{2{m}_{i}})+ \sum_{i<j=1}
^{6}V_{ij}-T_{CM} \nonumber \\
V_{ij} &=& V_{ij}^{c}+V_{ij}^{oge} \nonumber \\ 
V_{ij}^{oge}&=&{\alpha}_{s}\frac{1}{4}{\lambda}_i^c\cdot{\lambda}_j^c
[\frac{1}{r}
-\frac{\pi}{2}\delta(r_{ij})(\frac{1}{m_i^2}+\frac{1}{m_j^2}+
\frac{4{\sigma}_i{\sigma}_j}{3m_im_j})+\cdot\cdot\cdot] \nonumber
\end{eqnarray}
\[~~~~~~{V}_{ij}^{c}=\left\{\begin{array}{cc} 
-{\lambda}_{i}^{c} {\lambda}_{j}^{c}a_c{r}_{ij}^{2} &
\mbox{if i,j occur in the same baryon orbit} \\
-{\lambda}_{i}^{c}{\lambda}_{j}^{c}\frac{a_c}{\mu}
(1-{e}^{-{\mu}{r}_{ij}^2})
& \mbox{if i,j occur in different baryon orbits} \\
\end{array}
\right. \]

and the delocalized quark orbits are
\begin{eqnarray}
{\Psi}_{L}(r) &=& \frac{({\Phi}_{L}(r)+\epsilon({R}_{s}){\Phi}_{R}(r))}{N({R}_{s})} 
\nonumber \\
{\Psi}_{R}(r) &=& \frac{({\Phi}_{R}(r)+\epsilon({R}_{s}){\Phi}_{L}(r))}{N({R}_{s})} 
\nonumber \\
{N}^{2}({R}_{s}) &=& 1+{\epsilon}^{2}({R}_{s})+2{\epsilon}({R}_{s})
\langle{\Phi}_{L}|{\Phi}_{R}\rangle \nonumber 
\end{eqnarray}
where ${\phi}_L ({\phi}_R)$ is the left- (right-) centered quark orbital
wave function; the mixing parameter $\epsilon(R_s)$ is determined
variationally for every $R_s$ by the dynamics of the six-quark system.
\[m_u=m_d=313MeV, m_s=634MeV,\]
\[b=0.603fm, {\alpha}_s=1.54, a_c=25.13MeV{fm}^{-2},
\mu=1.0fm^{-2}. \]

Among these three models, the QDCSM has the fewest parameters.

In order to do a systematic calculation for all three-flavor (u, d, s)
six-quark systems, the adiabatic approximation is used to obtain the
effective potential between two baryons as a function of their
separation. The mass of the six-quark system is then estimated as the
sum of the two separate baryon masses, plus the value of the effective
potential at its minimum and the zero-point oscillation energy.  In the
calculation of the six-body matrix elements, the fractional parentage
expansion is used. The details of this method can be found in
Refs.\cite{wpwg,jlping}.  Such an adiabatic calculation is reasonable
for a systematic survey, but only gives a rough estimation of the mass
of the six-quark system. A dynamical calculation is needed to obtain
quantitative results, but is not expected to lead to significant 
differences at the level of accuracy of interest here.

\section{Results and Discussion}
 The baryon-baryon scattering properties calculated are listed in Tables
I to VII.  Figures 1-12 show the baryon-baryon effective potentials (in
units of MeV) as a function of separation $R$(fm).

\subsection{Two smearing versions of the GBE model}

In a series of papers, Glozman et al. used two different versions of
smearing of Goldstone boson exchanges to calculate baryon masses and
the NN interaction. We have extended their studies to all possible
baryon-baryon S-wave channels consisting of baryon octet and decuplet 
states and find:

(1) The short-range repulsion obtained from Gaussian smearing is much
larger than from Yukawa smearing for some states. Taking the (001)
deuteron channel as an example, we find the short-range repulsion is
2436 MeV for Gaussian smearing\cite{light}, but only about 790MeV for
Yukawa smearing.

(2) Channel coupling has obvious effects for model(a), but negligible
effect for model(b). Because of this, some states (about 1/6 of the
total) have the similar effective interactions in models (a) and (b)
for single channel results, but quite different effective interactions
for results from channel coupling. Such states include (-1,3/2,1)
(-1,3/2,2) (-201) (-200) (-213) (-221) (-3,1/2,2) (-3,3/2,3) (-402)
(-412) , typical examples of which are shown in Fig. 1.

(3) We identify three groups of states, depending on the
minimum value of the potential, $|V_0|$:

For the first group, the difference of $|V_0|$ between models (a) and
(b) is about 10 MeV. These states include most of the states with small
angular momentum ($J=0,1$).  These states, including (021),
(030), (-1,1/2,2),  (-1,3/2,0), (-1,5/2,0),
(-2,0,2), (-2,2,0), (-5,1/2,0), (-6,0,2) and so
on, are not sensitive to the different forms of smearing of the
$\delta$-function, typical examples of which are shown in Fig. 2.

For the second group, the difference of $|V_0|$ values between model(a)
and (b) is greater than 20 MeV, but not more than 100 MeV.  For these
states, intermediate-range attraction is always present, but the
strength of the attraction varies with the form of smearing.  Many
states with large angular momentum ($J=3$) belong to this group, such
as (003), (013), (-203), (-3,1/2,0), (-3,1/2,3), (-403) and so on. (See
Fig. 2).  As for the first group, the results do not depend
sensitively on the form of the choice of smearing.

The difference between models (a) and (b) shows up most strongly for
the third group of states.  Comparing the results for (023), (022),
(033), (-1,3/2,3), (-1,5/2,3), (-1,5/2,2), (-213), (-3,3/2,0), (-3,3/2,3),
(-5,1/2,2), (-5,1/2,3), and (-603), (see Fig. 3,) we find a
very small value for $|V_0|$, or even no attraction at all, in model
(b) , while we find a very large value for $|V_0|$, in the range of 200 to
400 MeV at small equilibrium separation ($R_s = 0.5-0.8 fm$) in
model (a).

In the following comparisons, when we refer to the GBE model, we will
mean model (b), because it is consistent with the requirement that the
volume integral of the pseudoscalar meson exchange interaction should
vanish.

\subsection{Comparison among the three different models}

All possible S-wave states of (SIJ) obtained from the ground state
octet and decuplet baryons in the u, d, and s three-flavor world have
been calculated with the three models (GBE(a) and (b), FN and QDCSM)
based on three sets of effective degrees of freedom.  We will discuss
the general features first and then the details for some interesting
aspects.

\subsubsection{General features} 

(1) Deuteron properties: 
Since the deuteron is a well known, stable, two-nucleon state
($SIJ=001$), we take it as a measure of the precision of our model
calculation.  The three different models all show a weak attraction;
the minimum of the effective potential is {-18~MeV~(FN)}, -21 MeV (QDCSM),
-3 MeV (GBE), and the separation corresponding to the minimum energies is
1.2fm, 1.3fm, 2.0fm, respectively.  These reproduce the deuteron
channel NN interaction qualitatively, but the estimated masses in the
three models differ by 15-27 MeV from the deuteron energy.  The main
causes of this deviation are twofold:  First, the S-D wave mixing due
to the one pion exchange tensor interaction has not been taken into
account; second the adiabatic approximation and zero-point harmonic
oscillation energy used in this calculation are rough approximations.
A dynamical calculation with the extended QDCSM, where the one pion
exchange tensor interaction is included, does reproduce the deuteron
energy, size and  D-wave mixing accurately\cite{wpwg}.  This emphasizes 
that the results reported here only produce a qualitative effective
baryon-baryon interaction. To study dibaryon states, a careful
dynamical calculation is needed.  In fact, if the deuteron would not
have been found by experiment, it would have been difficult for any of
the models to predict its existence.

(2) Quark delocalization effect:
For Fujiwara's quark-meson-gluon coupling model, enlarging the model
space will not change the results, i.e., systems always prefer to
remain with small delocalization($\epsilon\leq{0.1}$)
\footnote{In Tables III--VII, $\epsilon$ is omitted for the 
FN model.}, even
for those states with minimum energy at small separations.  A similar
conclusion follows for both of Glozman's quark-meson coupling models,
(a) and (b).  On the other hand, for the QDCSM, the quark
delocalization effect is obvious. For some states the delocalization
parameter $\epsilon$ can reach 1.0 at the minimum energy of the system.
Quark delocalization compensates for the explicit meson exchange effect
absent in the QDCSM.

(3) Channel coupling effect:  In all of these three models, the channel
coupling effect is small for most cases; the largest effect is about
50-100MeV. For those channels where the binding energy is close to the
channel threshold, channel coupling may well play a critical role in
forming dibaryon states. Jaffe's H particle case is a clear example. In
those cases, the channel coupling should be studied dynamically.

\subsubsection{Details} 

From the calculated results of three different models, we found that
about two-thirds of the states within the u, d,and s three-flavor world
have quite similar, and in some cases indistinguishable, effective
baryon-baryon interactions:

(1) States with the same, purely repulsive, interaction: \\ 
States with almost the same effective repulsive baryon-baryon
interactions include (000), (011), (023), (022), (033), (032), 
(-1,3/2,2), (-1,3/2,1), (-1,5/2,3), (-1,5/2,2), (-221), (-403), 
(-411), (-5,1/2,3), (-5,1/2,2), (-603) and (-602). 
These states have purely repulsive potentials in all three models
despite their different effective degrees of freedom.  There are
altogether 17 states, among all 64 states of 6-quark system, belonging
to this case.  To simplify the presentation, only the four states which
have different strangeness are shown in Fig.4.

(2) States with similar weak attraction :\\
{\indent}{\indent}(i) Some states have very similar baryon-baryon
interactions for all three models, all producing a weak attraction,
with the minimum energies of the systems appearing at large separation.
(The separation is a bit smaller for a few states in the FN model).  We
find 13 states belonging to this case, such as (001), (010), (021),
(-1,1/2,1), (-1,3/2,0), (-201), (-211), (-210), (-222), (-3,3/2,2), 
(-3,3/2,1), (-402) and (-400). Typical ones are shown in Fig.5.
Because the binding energies of these states are small, their masses
are close to the corresponding thresholds. Whether the binding energy
is positive or negative is very sensitive to the details of the
models.  It is difficult to determine whether these are strong
interaction stable states or not, since the precision of the adiabatic
approximation is limited.  A dynamical calculation is especially needed
in these cases in order to obtain more reliable results.\\
{\indent}{\indent}(ii) There are states for which the QDCSM and FN give
quite similar results, but the GBE model yields an even weaker
attraction.  States belonging to this case include (031), (-3,1/2,1),
(-401) and (-412), two of which are shown in Fig.6. The QDCSM and FN model
obtain a minimum of the effective attraction of several tens of MeV, or
at most a hundred MeV, but the GBE model yields only several MeV for
these states.\\
{\indent}{\indent}(iii) There are states for which the QDCSM and GBE
models obtain a weak attraction, while the FN model yields a little
stronger attraction. These states include: (-1,1/2,0), (-1,5/2,1),
(-220) and (-3,1/2,0), as shown in Fig.7. For these states, the QDCSM
and GBE models obtain a minimum of several MeV, while in the FN model,
the minimum is several tens of MeV.\\
{\indent}{\indent}(iv) For some states, the QDCSM obtains a somewhat
stronger attraction than either of the other two models.  Such states
include:  (012), (-223) and (-413); the latter two are shown in Fig.8.\\
{\indent}{\indent}(v) Finally, there are states for which the FN and
GBE models obtain slightly different effective potentials with the
results of the QDCSM in between. For the states (020), (030) and
(-1,5/2,0), the latter two of which are shown in Fig.9, the GBE model
yields a strong repulsive core and weak intermediate-range attraction; the
FN model obtains a weak repulsive core and considerable intermediate-range
attraction, and the QDCSM obtains, on the one hand, a repulsive core
similar to that of the GBE model, and on the other, an
intermediate-range attraction similar to that of the FN model.

Altogether, we find that the three different models based on the three
different effective degrees of freedom not only all give a good
description of baryon spectra, but they also give similar effective
baryon-baryon interactions for two thirds of all of the states possible
within the three-flavor world. This implies that, with regard to baryon
spectra and baryon-baryon interactions, the models based on the
different effective degree of freedoms are still compatible. We infer
that quark delocalization and color screening, working together in the
QDCSM, must provide a good representation of meson exchange effects at
some level of accuracy.  All of this appears consistent with old
conclusions developed from nonperturbative QCD:  Models with different
effective degrees of freedom but with physically equivalent results can
be obtained for low energy QCD under different approximations.

Each model has distinct characteristics with respect to the remaining
one third of the states:

(1) The QDCSM predicts strong intermediate-range attraction for states
with large angular momentum. As the strangeness becomes more negative,
the minimum of the effective potential, $V_0$, becomes less negative,
as can be seen from the sequence: (003) with $V_0=-359$MeV, (-1,1/2,3)
with $V_0=-308$MeV, (-202) with $V_0=-184$MeV, and (-403) where $|V_0|$
is only several MeV. (See Fig.10.)

(2) The FN model predicts a weak attraction in the effective potential
for states with small strangeness. Conversely to the QDCSM, here, 
as the
strangeness becomes more negative, $|V_0|$ also becomes more negative.
For example, for the state (-3,3/2,0) $V_0=-87$MeV, for (-410)
$V_0=-95$MeV, for (-5,1/2,1) $V_0=-136$MeV, and for (-600)
$V_0=-281$MeV. (See Fig.11.)

The Darwin term of the Breit-Fermi interaction has been included in the
FN model.  We find its contribution to the baryon-baryon effective
interaction to be small, typically giving rise to a weak attraction
with a minimum ${\leq} 10 MeV$.

(3) Since the GBE model that we used in our calculations does not take
into account scalar and vector meson contributions, the effective
interaction obtained is usually less attractive than the other two
models.  One would expect that the GBE will give more similar results
if these exchanges are incorporated.

\subsubsection{Particularly interesting states}

(1) $d^{*}$ dibaryon state (003); (see Fig.10): The QDCSM predicts this
to be a tightly bound, 6-quark state, but the other models predict only
a weak intermediate-range attraction, which does not provide enough
attraction to form a bound state relative to $NN\pi\pi$ threshold.
Recently, a dynamical calculation in the QDCSM obtained a slightly
smaller binding energy for the $d^{*}$ ($\sim$ 2.18GeV)\cite{wpwg}.
The difference is due to the fact that the zero-point oscillation
energy in this case underestimates the relative kinetic energy of the
two baryons.  However the $d^{*}$ mass is still much lower than found
in the other models.

(2) Di-$\Omega$ state (-600); (see Fig.11): The FN model predicts a
strong intermediate-range attraction with binding energy $B=-62$MeV.
The mass of the state is 3282 MeV, which is lower than the
$\Omega\Omega$ threshold $3345 MeV$, so that a strong interaction
stable state is predicted.  Another hybrid model\cite{zhang} obtains a
similar result; (B is about -100 MeV). The GBE model predicts $B=-11$
MeV. This mass is slightly lower than the threshold, and is expected to
decrease further if scalar and vector meson exchanges are incorporated.
The QDCSM predicts a mass of 3350 MeV, but a dynamical calculation
obtains a smaller mass, which is closer to the results of the hybrid
models.  In this case, the zero-point harmonic oscillation energy
overestimates the relative kinetic energy.

(3) H particle (-200); (see Fig.12): The QDCSM predicts that this is a
strong interaction stable state with small binding energy, $B=-14$ MeV;
the mass is 2218 MeV. The FN model obtains a mass of 2312 MeV, which
is higher than the threshold (2231MeV).  Zhang's dynamical calculation
obtains a mass for the H close to the $\Lambda\Lambda$
threshold\cite{zhangg}.  The GBE(a) model obtains 2207 MeV, which is a
little lower than the threshold, while GBE(b) obtains 2249 MeV, which
is a little higher than the threshold.  Almost all of these model
results are close to the $\Lambda\Lambda$ threshold.  Taken together,
they imply that the mass of the H particle is sensitive to model
details. Experimental searches should (and do) take into account the
two possibilities of a strong interaction stable state with long
lifetime, as well as a narrow resonances with a short lifetime.

\section{conclusion}
The three constituent quark models based on quark-meson, quark-gluon
and quark-meson-gluon effective degrees of freedom are widely used in
the study of hadron spectroscopy and hadron interaction, and there is a
hot debate about which degrees of freedom are the proper
ones\cite{twenone}. For the $64$ two baryon channels consisting of
octet and decuplet baryons, we find that the three models give similar
effective baryon-baryon interactions for about 2/3 of these channels.
The three models appear to be different, but in fact are not very
different in describing effective baryon-baryon interactions.

For some time, a number of authors have viewed the QDCSM as an
exceptional model. Some have even claimed that it violates principles
of quantum mechanics\cite{zzhang}. This is incorrect, as it is a
conventional effective matrix element approach, which we have proposed
because of our concern regarding a sound theoretical basis for direct
extension of the two body confinement interaction from single to
multi-hadron systems, where nonlinearity may be expected to be
significant.  Quark model studies of multi-hadron systems provide
important checks of the model descriptions of confinement. The results
of this paper demonstrate that predictions of the QDCSM are not so
exceptional as might appear from its definition.  On the contrary, it
embodies meson exchange effects in its own way, i.e., through quark
delocalization and a different parametrization of quark confinement.
Since the effective matrix element method is widely used in the
Heisenberg version of matrix mechanics, we are at a loss to explain 
how the authors of Ref.\cite{zzhang} have so misunderstood
our model as to arrive at such incorrect conclusions regarding it.

This paper also demonstrates that there are differences among these
three models.  This has been shown in baryon
spectroscopy\cite{twenone}.  It also appears in the predictions for
different dibaryon masses, especially for states of high spin and high
strangeness.  More precise hadron spectroscopy and experimental study
of the dibaryon states predicted by these models are needed to
distinguish which degree of freedom best describes low energy hadron
physics.

This research is partly supported by the NSF, SED and SSTD of China,
and partly by the US Department of Energy under contract
W-7405-ENG-36.

\newpage

Table I: An asymptotic two-baryon system with channel quantum
numbers and delocalization parameter values ($\epsilon$) in GBE(a) and (b)
for $S=0,-1,-2$. The notation is as follows: sc stands for single channel; cc
stands for multichannel coupling; $B_{\alpha}$ is the binding energy;
$E_{\alpha}$ is the mass and $V_0$ is the potential energy at the
equilibrium separation, $R_s$; (SIJ) are the strangeness, isospin and
total angular momentum quantum numbers for each channel.  

\begin{tabular}{c|c||c|c|c|c|c||c|c|c|c|c||c} \hline
\multicolumn{2}{c||}{Table I }&\multicolumn{5}{c||}{GBE(a)}&\multicolumn{5}{c||}{GBE(b)} \\ \hline 
SIJ&&$E_{\alpha}$&$V_0$&$B_{\alpha}$&$\epsilon$&$R_s$ 
&$E_{\alpha}$&$V_0$&$B_{\alpha}$&$\epsilon$&$R_s$&Threshold \\ \hline
$0,0,1$&sc
&$1889$&$-1$&$11$&$0.0$&$2.3$ 
&$1891$&$-3$&$13$&$
0.0$&$2.0$&$1878(NN)$ \\ \cline{1-12}
&cc
&$1889$&$-1$&$11$&$0.0$&$2.3$ 
&$1891$&$-3$&$13$&$
0.0$&$2.0$& \\ \hline
$0,0,3$&sc
&$2503$&$-35$&$39$&$0.1$&$0.7$ 
&$2474$&$-14$&$10$&$0.1$&$1.4$&$2464(\Delta\Delta)$,$2158(NN\pi\pi)$\\ \hline
$0,1,0$&sc
&$1888$&$-1$&$10$&$0.0$&$2.4$ 
&$1892$&$-3$&$14$&$0.0$&$1.9$&$1878(NN)$ \\ \cline{1-12}
&cc
&$1888$&$-1$&$10$&$0.0$&$2.4$ 
&$1892$&$-3$&$14$&$0.0$&$1.9$& \\ \hline
$0,2,3$&sc
&$2333$&$-205$&$-131$&$0.0$&$0.8$ 
&$2471$&$-1$&$7$&$0.0$&$2.6$&$2464(\Delta\Delta)$,$2158(NN\pi\pi)$\\ \hline\hline
$-1,1/2,3$&sc
&$2621$&$-40$&$5$&$0.1$&$1.0$ 
&$2619$&$-24$&$2$&$0.1$&$1.3$&$2617(\Delta{\Sigma}^{*})$ \\ \cline{1-12}
&cc
&$2638$&$-49$&$21$&$0.1$&$0.8$ 
&$2619$&$-24$&$2$&$0.1$&$1.3$&$2335(N\Lambda\pi\pi)$ \\ \hline
$-1,3/2,0$&sc
&$2141$&$-1$&$9$&$0.0$&$2.4$ 
&$2144$&$-3$&$12$&$0.0$&$1.9$&$2132(N\Sigma)$ \\ \cline{1-12}
&cc
&$2141$&$-1$&$9$&$0.0$&$2.4$ 
&$2144$&$-3$&$12$&$0.0$&$1.9$& \\ \hline
$-1,3/2,3$&sc
&$2523$&$-164$&$-94$&$0.0$&$0.8$ 
&$2623$&$-1$&$6$&$0.0$&$2.5$&$2617(\Delta{\Sigma}^{*})$ \\ \cline{1-12}
&cc
&$2523$&$-164$&$-94$&$0.0$&$0.8$ 
&$2629$&$-10$&$13$&$0.1$&$1.4$&$2335(N\Lambda\pi\pi)$ \\ \hline\hline
$-2,0,0$&sc
&$2239$&$0$&$8$&$0.0$&$-$ 
&$2245$&$0$&$14$&$0.0$&$-$&$2231(\Lambda\Lambda)$ \\ \cline{1-12}
&cc
&$2207$&$-106$&$-24$&$0.0$&$0.8$ 
&$2249$&$-1$&$18$&$0.0$&$1.7$& \\ \hline
$-2,0,2$&sc
&$2481$&$-1$&$9$&$0.0$&$2.3$ 
&$2487$&$-3$&$15$&$0.0$&$1.7$&$2472(N{\Xi}^{*})$ \\ \cline{1-12}
&cc
&$2481$&$-1$&$9$&$0.0$&$2.3$ 
&$2487$&$-3$&$15$&$0.0$&$1.7$&$2397(N\Xi\pi)$ \\ \hline
$-2,0,3$&sc
&$2785$&$-26$&$16$&$0.1$&$1.0$ 
&$2779$&$-3$&$10$&$0.0$&$1.8$&$2770({\Sigma}^{*}{\Sigma}^{*})$ \\ \hline
$-2,1,3$&sc
&$2762$&$-56$&$-3$&$0.1$&$0.9$ 
&$2775$&$-16$&$10$&$0.1$&$1.3$&$2765(\Delta{\Xi}^{*})$ \\ \cline{1-12}
&cc
&$2701$&$-117$&$-64$&$0.0$&$0.9$ 
&$2765$&$-36$&$-1$&$0.1$&$1.1$&$2690(\Delta\Xi\pi)$ \\ \hline
$-2,2,0$&sc
&$2393$&$-1$&$7$&$0.0$&$2.5$ 
&$2397$&$-3$&$11$&$0.0$&$1.9$&$2386(\Sigma\Sigma)$ \\ \cline{1-12}
&cc
&$2393$&$-1$&$7$&$0.0$&$2.5$ 
&$2400$&$-3$&$14$&$0.0$&$1.7$& \\ \hline
$-2,2,2$&sc
&$2559$&$0$&$9$&$0.0$&$-$ 
&$2560$&$-1$&$10$&$0.0$&$2.1$&$2550(\Xi\Delta)$ \\ \cline{1-12}
&cc
&$2262$&$-471$&$-288$&$0.0$&$0.5$ 
&$2568$&$-9$&$18$&$0.1$&$1.3$& \\ \hline
$-2,2,3$&sc
&$2635$&$-197$&$-131$&$0.0$&$0.8$ 
&$2766$&$0$&$1$&$0.0$&$-$&$2765(\Delta{\Xi}^{*})$ \\ \cline{1-12}
&cc
&$2491$&$-393$&$-274$&$0.0$&$0.6$ 
&$2769$&$-22$&$3$&$0.1$&$1.3$&$2690(\Delta\Xi\pi)$ \\ \hline
\end{tabular}
\newpage
Table II: The same as Table I for states with $S=-3,-4,-5,-6$. 

\begin{tabular}{c|c||c|c|c|c|c||c|c|c|c|c||c} \hline
\multicolumn{2}{c||}{Table II }&\multicolumn{5}{c||}{GBE(a)}&\multicolumn{5}{c||}{GBE(b)} \\ \hline 
SIJ&&$E_{\alpha}$&$V_0$&$B_{\alpha}$&$\epsilon$&$R_s$ 
&$E_{\alpha}$&$V_0$&$B_{\alpha}$&$\epsilon$&$R_s$&Threshold \\ \hline
$-3,1/2,0$&sc
&$2440$&$0$&$7$&$0.0$&$-$ 
&$2449$&$-1$&$15$&$0.0$&$1.7$&$2434(\Lambda\Xi)$ \\ \cline{1-12}
&cc
&$2444$&$-50$&$10$&$0.0$&$0.9$ 
&$2455$&$-3$&$22$&$0.0$&$1.4$& \\ \hline
$-3,1/2,3$&sc
&$2867$&$-100$&$-51$&$0.0$&$0.9$ 
&$2924$&$-1$&$6$&$0.0$&$2.4$&$2918({\Xi}^{*}{\Sigma}^{*})$ \\ \cline{1-12}
&cc
&$2867$&$-100$&$-51$&$0.0$&$0.9$ 
&$2930$&$-9$&$12$&$0.1$&$1.4$& \\ \hline
$-3,3/2,0$&sc
&$2361$&$-197$&$-150$&$0.0$&$1.0$ 
&$2523$&$-7$&$21$&$0.1$&$1.3$&$2511(\Sigma\Xi)$ \\ \cline{1-12}
&cc
&$2290$&$-513$&$-221$&$0.0$&$0.4$ 
&$2534$&$-10$&$23$&$0.1$&$1.2$& \\ \hline
$-3,3/2,3$&sc
&$2874$&$-81$&$-30$&$0.1$&$0.9$ 
&$2917$&$-11$&$13$&$0.1$&$1.3$&$2904(\Delta\Omega)$ \\ \cline{1-12}
&cc
&$2701$&$-318$&$-203$&$0.0$&$0.6$ 
&$2892$&$-54$&$-12$&$0.1$&$1.0$&$2714(\Lambda\Xi\pi\pi)$ \\ \hline\hline
$-4,0,0$&sc
&$2637$&$0$&$1$&$0.0$&$-$ 
&$2654$&$-5$&$18$&$0.0$&$1.4$&$2636(\Xi\Xi)$ \\ \cline{1-12}
&cc
&$2637$&$0$&$1$&$0.0$&$-$ 
&$2654$&$-5$&$18$&$0.0$&$1.4$& \\ \hline
$-4,0,3$&sc
&$3038$&$-76$&$-29$&$0.0$&$0.9$ 
&$3073$&$-1$&$6$&$0.0$&$2.4$&$3066({\Xi}^{*}{\Xi}^{*})$ \\ \hline
$-4,1,2$&sc
&$2828$&$-52$&$-23$&$0.0$&$1.2$ 
&$2852$&$0$&$1$&$0.0$&$-$&$2851(\Xi{\Xi}^{*})$ \\ \cline{1-12}
&cc
&$2657$&$-360$&$-195$&$0.0$&$0.5$ 
&$2873$&$-13$&$21$&$0.1$&$1.1$& \\ \hline
$-4,1,3$&sc
&$2998$&$-107$&$-59$&$0.0$&$0.9$ 
&$3063$&$0$&$6$&$0.0$&$-$&$3057({\Sigma}^{*}\Omega)$ \\ \cline{1-12}
&cc
&$2876$&$-288$&$-181$&$0.0$&$0.6$ 
&$3068$&$-12$&$11$&$0.1$&$1.3$& \\ \hline\hline
$-5,1/2,2$&sc
&$2907$&$-123$&$-83$&$0.0$&$1.0$ 
&$2991$&$0$&$1$&$0.0$&$-$&$2990(\Xi\Omega)$ \\ \cline{1-12}
&cc
&$2778$&$-322$&$-212$&$0.0$&$0.6$ 
&$2998$&$0$&$7$&$0.0$&$-$& \\ \hline
$-5,1/2,3$&sc
&$3025$&$-256$&$-181$&$0.0$&$0.7$ 
&$3207$&$0$&$1$&$0.0$&$-$&$3205({\Xi}^{*}\Omega)$ \\ \cline{1-12}
&cc
&$3025$&$-256$&$-181$&$0.0$&$0.7$ 
&$3213$&$0$&$7$&$0.0$&$-$& \\ \hline\hline
$-6,0,0$&sc
&$3350$&$0$&$5$&$0.0$&$2.6$ 
&$3334$&$-40$&$-11$&$0.1$&$1.1$&$3345(\Omega\Omega)$ \\ \cline{1-12}
$-6,0,1$&sc
&$3350$&$0$&$5$&$0.0$&$2.6$ 
&$3355$&$-11$&$10$&$0.1$&$1.3$& \\ \cline{1-12}
$-6,0,3$&sc
&$3198$&$-218$&$-147$&$0.0$&$0.7$ 
&$3346$&$0$&$1$&$0.0$&$-$& \\ \hline
\end{tabular}
\newpage
Table III: An asymptotic two-baryon system with channel quantum numbers
and delocalization parameter values ($\epsilon$) in the QDCSM, GBE and
FN models for $S=0$. The notation is the same as Table I.  (For the FN 
model, $\epsilon$ is omitted.)

\begin{tabular}{c|c||c|c|c|c|c||c|c|c|c||c|c|c|c|c||c} \hline
\multicolumn{2}{c||}{Table III}&
\multicolumn{5}{c||}{GBE(b)}& 
\multicolumn{4}{c||}{FN}&\multicolumn{5}{c||}{QDCSM} \\ \hline 
SIJ&&$E_{\alpha}$&$V_0$&$B_{\alpha}$&$\epsilon$&$R_s$&$E_{\alpha}$&$V_0$&$B_{\alpha}$&$R_s$&$E_{\alpha}$&$V_0$&$B_{\alpha}$&$\epsilon$&$R_s$&Threshold \\ \hline
$0,0,0$&sc&$1879$&$0$&$1$&$0.0$&-&$1879$&$0$&$1$&-&$1879$&$0$&$1$&$0.0$&-&$1878(NN)$\\ \cline{1-16}
&cc&$1888$&$4$&$10$&$0.0$&-&$1887$&$2$&$9$&-&$1885$&$0$&$7$&$0.0$&-& \\ \cline{1-16}
$0,0,1$&sc&$1891$&$-3$&$13$&$0.0$&$2.0$&
$1903$&$-18$&$25$&$1.2$&$1885$&$-20$&$7$&$0.1$&$1.5$& \\ \cline{1-16}
&cc&$1891$&$-3$&$13$&$0.0$&$2.0$&
$1903$&$-18$&$25$&$1.2$&$1894$&$-21$&$16$&$0.2$&$1.3$& \\ \hline
$0,0,2$&sc&$2478$&$-14$&$14$&$0.1$&$1.3$&
$2479$&$-24$&$15$&$1.1$&$2259$&$-233$&$-205$&$1.0$&$1.3$&$2464(\Delta\Delta)$ 
\\ \cline{1-16}
$0,0,3$&sc&$2474$&$-14$&$10$&$0.1$&$1.4$&
$2479$&$-24$&$15$&$1.1$&$2144$&$-359$&$-320$&$1.0$&$1.1$&$2158(NN\pi\pi)$ 
\\ \hline 
$0,1,0$&sc&$1892$&$-3$&$14$&$0.0$&$1.9$&
$1906$&$-35$&$28$&$1.0$&$1893$&$-10$&$15$&$0.1$&$1.6$&$1878(NN)$ \\ \cline{1-16}
&cc&$1892$&$-3$&$14$&$0.0$&$1.9$&
$1905$&$-35$&$27$&$1.0$&$1892$&$-9$&$14$&$0.1$&$1.6$& \\ \cline{1-16}
$0,1,1$&sc&$1879$&$0$&$1$&$0.0$&$-$&
$1879$&$0$&$1$&$-$&$1879$&$0$&$1$&$0.0$&-& \\ \cline{1-16}
&cc&$1885$&$0$&$7$&$0.0$&-&
$1890$&$0$&$12$&$0.0$&$1885$&$0$&$7$&$0.0$&-& \\ \hline
$0,1,2$&sc&$2182$&$-2$&$11$&$0.0$&$2.1$&
$2185$&$-14$&$14$&$1.4$&$2122$&$-87$&$-49$&$0.4$&$1.2$&$2171(N\Delta)$ \\ \cline
{1-16}
&cc&$2181$&$-4$&$11$&$0.0$&$1.9$&
$2185$&$-14$&$14$&$1.4$&$2122$&$-87$&$-49$&$0.4$&$1.2$&$2018(NN\pi)$ \\ \hline
$0,1,3$&sc&$2472$&$-4$&$8$&$0.0$&$2.0$&
$2476$&$-3$&$12$&$1.8$&$2304$&$-188$&$-160$&$1.0$&$1.3$&$2464(\Delta\Delta)$ \\ \cline{1-16}
$0,2,0$&sc&$2476$&$-12$&$12$&$0.1$&$1.4$&$2442$&$-80$&$-22$&$0.9$&$2383$&$
-105$&$-81$&$0.4$&$1.4$& \\ \hline
$0,2,1$&sc&$2181$&$-5$&$10$&$0.0$&$1.9$&
$2170$&$-39$&$1$&$1.2$&$2169$&$-24$&$-2$&$0.1$&$1.6$&$2171(N\Delta)$ \\ \cline{1-16}
&cc&$2181$&$-5$&$10$&$0.0$&$1.9$&
$2186$&$-40$&$15$&$1.0$&$2169$&$-24$&$-2$&$0.1$&$1.6$&$2018(NN\pi)$ \\ \cline{1-16}
$0,2,2$&sc&$2172$&$0$&$1$&$0.0$&$-$&
$2172$&$0$&$1$&$-$&$2172$&$0$&$1$&$0.0$&$-$& \\ \cline{1-16}
&cc&$2180$&$-1$&$9$&$0.0$&$-$&
$2172$&$0$&$6$&$-$&$2172$&$0$&$6$&$0.0$&$-$& \\ \hline
$0,2,3$&sc&$2471$&$-1$&$7$&$0.0$&$-$&$2470$&$0$&$6$&$-$&$2442$&$-41$&$
-22$&$0.6$&$1.6$&$2464(\Delta\Delta)$ \\ \cline{1-16}
$0,3,0$&sc&$2471$&$-11$&$7$&$0.0$&$1.6$&
$2385$&$-153$&$-79$&$0.8$&$2416$&$-69$&$-48$&$0.2$&$1.5$& \\ \cline{1-16}
$0,3,1$&sc&$2473$&$-6$&$9$&$0.0$&$1.8$&
$2440$&$-63$&$-24$&$1.1$&$2451$&$-32$&$-13$&$0.2$&$1.6$& \\ \cline{1-16}
$0,3,2$&sc&$2472$&$-1$&$8$&$0.0$&$2.3$&
$2474$&$-3$&$10$&$1.9$&$2474$&$-2$&$10$&$0.1$&$2.0$& \\ \cline{1-16}
$0,3,3$&sc&$2465$&$0$&$1$&$0.0$&$-$&
$2465$&$0$&$1$&$-$&$2465$&$0$&$1$&$0.0$&$-$& \\ \hline
\end{tabular}
\newpage
Table IV: The same as Table III for $S=-1$.

\begin{tabular}{c|c||c|c|c|c|c||c|c|c|c||c|c|c|c|c||c} \hline
\multicolumn{2}{c||}{Table IV}&
\multicolumn{5}{c||}{GBE(b)}& 
\multicolumn{4}{c||}{FN}&\multicolumn{5}{c||}{QDCSM} \\ \hline 
SIJ&&$E_{\alpha}$&$V_0$&$B_{\alpha}$&$\epsilon$&$R_s$&$E_{\alpha}$&$V_0$&$B_{\alpha}$&$R_s$&$E_{\alpha}$&$V_0$&$B_{\alpha}$&$\epsilon$&$R_s$&Threshold \\ \hline
$-1,1/2,0$&sc&$2056$&$0$&$1$&$0.0$&$-$&$
2056$&$0$&$1$&$-$&$2056$&$0$&$1$&$0.0$&$-$&$2055(N\Lambda)$ \\ \cline{1-16}
&cc&$2070$&$-2$&$16$&$0.0$&$1.8$&$
2090$&$-35$&$35$&$0.9$&$2072$&$-5$&$17$&$0.1$&$1.6$& \\ \cline{1-16}
$-1,1/2,1$&sc&$2056$&$0$&$1$&$0.0$&$-$&$
2065$&$0$&$11$&$2.3$&$2056$&$0$&$1$&$0.0$&$-$& \\ \cline{1-16}
&cc&$2071$&$-1$&$17$&$0.0$&$1.8$&$
2095$&$-10$&$38$&$1.1$&$2077$&$-17$&$23$&$0.2$&$1.2$& \\ \hline
$-1,1/2,2$&sc&$2334$&$-2$&$11$&$0.0$&$2.0$&$
2339$&$-12$&$15$&$1.4$&$2255$&$-111$&$-68$&$0.4$&$1.1$&$2324(N{\Sigma}^{*})$ \\ \cline{1-16}
&cc&$2337$&$-3$&$13$&$0.0$&$1.8$&$
2359$&$-16$&$36$&$1.0$&$2227$&$-161$&$-96$&$1.0$&$0.9$& \\ \hline
$-1,1/2,3$&sc&$2619$&$-24$&$2$&$0.1$&$1.3$&$
2630$&$-4$&$14$&$1.6$&$2346$&$-308$&$-271$&$1.0$&$1.1$&$2617(\Delta{\Sigma}^{*})$ \\ \cline{1-16}
&cc&$2619$&$-24$&$2$&$0.1$&$1.3$&$
2631$&$-23$&$14$&$1.1$&$2346$&$-308$&$-271$&$1.0$&$1.1$&$2335(N\Lambda\pi\pi)$ \\ \hline
$-1,3/2,0$&sc&$2144$&$-3$&$12$&$0.0$&$1.9$&$
2143$&$-27$&$11$&$1.2$&$2138$&$-18$&$6$&$0.1$&$1.5$&$2132(N\Sigma)$ \\ \cline{1-16}
&cc&$2144$&$-3$&$12$&$0.0$&$1.9$&$
2156$&$-45$&$24$&$0.9$&$2138$&$-18$&$6$&$0.1$&$1.5$& \\ \cline{1-16}
$-1,3/2,1$&sc&$2133$&$0$&$1$&$0.0$&$-$&$
2133$&$0$&$1$&$-$&$2133$&$0$&$1$&$0.0$&$-$& \\ \cline{1-16}
&cc&$2133$&$0$&$1$&$0.0$&$-$&$
2133$&$0$&$1$&$-$&$2133$&$0$&$1$&$0.0$&$-$& \\ \hline
$-1,3/2,2$&sc&$2325$&$0$&$1$&$0.0$&$-$&$
2325$&$0$&$1$&$-$&$2325$&$0$&$1$&$0.0$&$-$&$2324(N{\Sigma}^{*})$ \\ \cline{1-16}
&cc&$2330$&$0$&$6$&$0.0$&$-$&$
2331$&$0$&$8$&$-$&$2333$&$-17$&$9$&$0.2$&$1.4$& \\ \hline
$-1,3/2,3$&sc&$2623$&$-1$&$6$&$0.0$&$-$&$
2622$&$0$&$5$&$-$&$2512$&$-131$&$-105$&$0.6$&$1.3$&$2617(\Delta{\Sigma}^{*})$ \\ \cline{1-16}
&cc&$2629$&$-10$&$13$&$0.1$&$1.4$&$
2628$&$-1$&$11$&$1.9$&$2512$&$-131$&$-105$&$0.6$&$1.3$&$2335(N\Lambda\pi\pi)$ \\ \cline{1-16}
$-1,5/2,0$&sc&$2621$&$-19$&$4$&$0.1$&$1.4$&$
2520$&$-167$&$-97$&$0.8$&$2578$&$-61$&$-38$&$0.2$&$1.4$& \\ \cline{1-16}
&cc&$2621$&$-19$&$4$&$0.1$&$1.4$&$
2520$&$-167$&$-97$&$0.8$&$2578$&$-61$&$-38$&$0.2$&$1.4$& \\ \hline
$-1,5/2,1$&sc&$2435$&$-2$&$10$&$0.0$&$2.0$&$
2420$&$-39$&$-5$&$1.2$&$2423$&$-21$&$-2$&$0.1$&$1.6$&$2425(\Delta\Sigma)$ \\ \cline{1-16}
&cc&$2437$&$-5$&$12$&$0.0$&$1.7$&$
2418$&$-66$&$-7$&$0.9$&$2424$&$-25$&$-1$&$0.2$&$1.4$& \\ \cline{1-16}
$-1,5/2,2$&sc&$2426$&$0$&$1$&$0.0$&$-$&$
2426$&$0$&$1$&$-$&$2426$&$0$&$1$&$0.0$&$-$& \\ \cline{1-16}
&cc&$2431$&$0$&$6$&$0.0$&$-$&$
2430$&$0$&$5$&$-$&$2430$&$0$&$5$&$0.0$&$-$& \\ \hline
$-1,5/2,3$&sc&$2618$&$0$&$1$&$0.0$&$-$&$2618$&$0$&$1$&$-$&$2681$&$0$&$2$&$0.0$&$-$&
$2617(\Delta{\Sigma}^{*})$ \\ \cline{1-16}
&cc&$2618$&$0$&$1$&$0.0$&$-$&$2618$&$0$&$1$&$-$&$2623$&$-7$&$6$&$0.2$&$1.8$&$
2335(N\Lambda\pi\pi)$ \\ \hline
\end{tabular}
\newpage
Table V: The same as Table III for $S=-2$.

\begin{tabular}{c|c||c|c|c|c|c||c|c|c|c||c|c|c|c|c||c} \hline
\multicolumn{2}{c||}{Table V}&
\multicolumn{5}{c||}{GBE(b)}& 
\multicolumn{4}{c||}{FN}&\multicolumn{5}{c||}{QDCSM} \\ \hline 
SIJ&&$E_{\alpha}$&$V_0$&$B_{\alpha}$&$\epsilon$&$R_s$&$E_{\alpha}$&$V_0$&$B_{\alpha}$&$R_s$&$E_{\alpha}$&$V_0$&$B_{\alpha}$&$\epsilon$&$R_s$&Threshold \\ \hline
$-2,0,0$&sc&$2245$&$0$&$14$&$0.0$&$1.9$&$
2302$&$-36$&$71$&$0.7$&$2249$&$-5$&$18$&$0.1$&$1.5$&$2231(\Lambda\Lambda)$ \\ \cline{1-16}
&cc&$2249$&$-1$&$18$&$0.0$&$1.7$&$
2312$&$-65$&$81$&$0.6$&$2218$&$-159$&$-14$&$1.0$&$0.6$& \\ \cline{1-16}
$-2,0,1$&sc&$2232$&$0$&$1$&$0.0$&$-$&$
2245$&$-1$&$14$&$1.9$&$2232$&$0$&$1$&$0.0$&$-$&\\ \cline{1-16}
&cc&$2237$&$0$&$6$&$0.0$&$-$&$
2329$&$-9$&$98$&$0.7$&$2300$&$-38$&$69$&$1.0$&$0.7$& \\ \hline
$-2,0,2$&sc&$2487$&$-3$&$15$&$0.0$&$1.7$&$
2488$&$-10$&$16$&$1.4$&$2476$&$-173$&$-102$&$0.7$&$0.8$&$2472(N{\Xi}^{*})$ \\ \cline{1-16}
&cc&$2487$&$-3$&$15$&$0.0$&$1.7$&$
2515$&$-35$&$43$&$0.8$&$2367$&$-184$&$-106$&$1.0$&$0.8$&$2397(N\Xi\pi)$ \\ \hline
$-2,0,3$&sc&$2779$&$-3$&$10$&$0.0$&$1.8$&$
2782$&$-6$&$13$&$1.5$&$2656$&$-142$&$-113$&$0.7$&$1.2$&$2770({\Sigma}^{*}{\Sigma}^{*})$ \\ \hline
$-2,1,0$&sc&$2258$&$0$&$1$&$0.0$&$-$&$
2258$&$0$&$1$&$-$&$2258$&$0$&$1$&$0.0$&$-$&$2257(N\Xi)$ \\ \cline{1-16}
&cc&$2253$&$0$&$6$&$0.0$&$-$&$
2310$&$-31$&$53$&$0.8$&$2277$&$-7$&$20$&$0.1$&$1.4$& \\ \cline{1-16}
$-2,1,1$&sc&$2267$&$0$&$10$&$0.0$&$2.3$&$
2274$&$-2$&$16$&$1.7$&$2276$&$-5$&$19$&$0.1$&$1.5$&\\ \cline{1-16}
&cc&$2268$&$0$&$11$&$0.0$&$2.2$&$
2322$&$-19$&$65$&$0.8$&$2290$&$-33$&$33$&$0.4$&$0.9$& \\ \hline
$-2,1,2$&sc&$2473$&$0$&$1$&$0.0$&$-$&$
2473$&$0$&$1$&$-$&$2362$&$-260$&$-188$&$1.0$&$0.8$&$2472(N{\Xi}^{*})$\\ \cline{1-16}
&cc&$2478$&$0$&$6$&$0.0$&$-$&$
2480$&$0$&$8$&$-$&$2422$&$-129$&$-50$&$1.0$&$0.8$& \\ \hline
$-2,1,3$&sc&$2775$&$-16$&$10$&$0.1$&$1.3$&$
2781$&$-3$&$16$&$1.5$&$2560$&$-251$&$-209$&$1.0$&$1.0$&$2765(\Delta{\Xi}^{*})$ \\ \cline{1-16}
&cc&$2765$&$-36$&$-1$&$0.1$&$1.1$&$
2782$&$-18$&$17$&$1.1$&$2557$&$-251$&$-209$&$1.0$&$1.0$&$2690(\Delta\Xi\pi)$ \\ \hline
$-2,2,0$&sc&$2397$&$-3$&$11$&$0.0$&$1.9$&$
2405$&$-57$&$19$&$0.8$&$2397$&$-11$&$11$&$0.1$&$1.5$&$2386(\Sigma\Sigma)$ \\ \cline{1-16}
&cc&$2400$&$-3$&$14$&$0.0$&$1.7$&$
2400$&$-62$&$14$&$0.8$&$2397$&$-11$&$11$&$0.1$&$1.5$& \\ \cline{1-16}
$-2,2,1$&sc&$2387$&$0$&$1$&$0.0$&$-$&$
2387$&$0$&$1$&$-$&$2387$&$0$&$1$&$0.0$&$-$& \\ \cline{1-16}
&cc&$2387$&$0$&$1$&$0.0$&$-$&$
2387$&$0$&$1$&$-$&$2387$&$0$&$1$&$0.0$&$-$& \\ \hline
$-2,2,2$&sc&$2560$&$-1$&$10$&$0.0$&$2.1$&$
2561$&$-2$&$11$&$1.9$&$2531$&$-57$&$-19$&$0.5$&$1.1$&$2550(\Delta\Xi)$ \\ \cline{1-16}
&cc&$2568$&$-9$&$18$&$0.1$&$1.3$&$
2588$&$-19$&$38$&$0.9$&$2530$&$-52$&$-20$&$0.3$&$1.2$& \\ \hline
$-2,2,3$&sc&$2766$&$0$&$1$&$0.0$&$-$&$
2766$&$0$&$1$&$-$&$2713$&$-82$&$-52$&$0.6$&$1.2$&$2765(\Delta{\Xi}^{*})$ \\ \cline{1-16}
&cc&$2769$&$-22$&$3$&$0.1$&$1.3$&$
2766$&$0$&$1$&$-$&$2717$&$-74$&$-49$&$0.4$&$1.3$&$2690(\Delta\Xi\pi)$ \\ \hline
\end{tabular}
\newpage
Table VI: The same as Table III for $S=-3,-6$.

\begin{tabular}{c|c||c|c|c|c|c||c|c|c|c||c|c|c|c|c||c} \hline
\multicolumn{2}{c||}{Table VI}&
\multicolumn{5}{c||}{GBE(b)}& 
\multicolumn{4}{c||}{FN}&\multicolumn{5}{c||}{QDCSM} \\ \hline 
SIJ&&$E_{\alpha}$&$V_0$&$B_{\alpha}$&$\epsilon$&$R_s$&$E_{\alpha}$&$V_0$&$B_{\alpha}$&$R_s$&$E_{\alpha}$&$V_0$&$B_{\alpha}$&$\epsilon$&$R_s$&Threshold \\ \hline
$-3,1/2,0$&sc&$2449$&$-1$&$15$&$0.0$&$1.7$&$
2453$&$-21$&$19$&$1.1$&$2444$&$-19$&$10$&$0.1$&$1.3$&$2434(\Xi\Lambda)$ \\ \cline{1-16}
&cc&$2455$&$-3$&$22$&$0.0$&$1.4$&$
2469$&$-63$&$36$&$0.7$&$2443$&$-19$&$9$&$0.1$&$1.3$& \\ \cline{1-16}
$-3,1/2,1$&sc&$2435$&$0$&$1$&$0.0$&$-$&$
2444$&$0$&$10$&$-$&$2528$&$-99$&$94$&$1.0$&$0.5$& \\ \cline{1-16}
&cc&$2495$&$-37$&$61$&$0.1$&$0.7$&$
2485$&$-83$&$51$&$0.6$&$2520$&$-107$&$86$&$1.0$&$0.5$& \\ \hline
$-3,1/2,2$&sc&$2623$&$0$&$12$&$0.0$&$2.0$&$
2629$&$-1$&$18$&$1.6$&$2552$&$-271$&$-151$&$1.0$&$0.6$&$2611(N\Omega)$ \\ \cline{1-16}
&cc&$2656$&$-15$&$45$&$0.1$&$0.9$&$
2672$&$-38$&$61$&$0.7$&$2548$&$-198$&$-64$&$1.0$&$0.6$&$2574(\Lambda\Xi\pi)$ \\ \hline
$-3,1/2,3$&sc&$2924$&$-1$&$6$&$0.0$&$2.4$&$
2923$&$0$&$5$&$-$&$2836$&$-115$&$-82$&$0.6$&$1.1$&$2918({\Sigma}^{*}{\Xi}^{*})$ \\ \cline{1-16}
&cc&$2930$&$-9$&$12$&$0.1$&$1.4$&$
2932$&$-6$&$14$&$1.4$&$2836$&$-115$&$-82$&$0.6$&$1.1$& \\ \hline
$-3,3/2,0$&sc&$2523$&$-7$&$21$&$0.1$&$1.3$&$
2512$&$0$&$1$&$-$&$2512$&$0$&$1$&$0.0$&$-$&$2511(\Sigma\Xi)$ \\ \cline{1-16}
&cc&$2534$&$-10$&$23$&$0.1$&$1.2$&$
2519$&$-87$&$8$&$0.7$&$2527$&$-8$&$15$&$0.1$&$1.4$& \\ \cline{1-16}
$-3,3/2,1$&sc&$2514$&$0$&$3$&$0.0$&$-$&$
2514$&$0$&$3$&$-$&$2527$&$-23$&$16$&$0.2$&$1.1$& \\ \cline{1-16}
&cc&$2538$&$-6$&$27$&$0.1$&$1.2$&$
2558$&$-49$&$47$&$0.7$&$2527$&$-23$&$16$&$0.2$&$1.1$& \\ \hline
$-3,3/2,2$&sc&$2707$&$0$&$4$&$0.0$&$-$&$
2709$&$0$&$6$&$-$&$2699$&$-57$&$-3$&$0.6$&$0.9$&$2703(\Xi{\Sigma}^{*})$ \\ \cline{1-16}
&cc&$2713$&$-26$&$10$&$0.1$&$1.1$&$
2726$&$-45$&$23$&$0.8$&$2683$&$-63$&$-19$&$0.4$&$1.0$&$2651(\Xi\Sigma\pi)$ \\ \hline
$-3,3/2,3$&sc&$2917$&$-11$&$13$&$0.1$&$1.3$&$
2913$&$0$&$8$&$-$&$2766$&$-202$&$-152$&$1.0$&$0.9$&$2904(\Delta\Omega)$ \\ \cline{1-16}
&cc&$2892$&$-54$&$-12$&$0.1$&$1.0$&$
2934$&$-12$&$29$&$1.0$&$2754$&$-201$&$-150$&$1.0$&$0.9$&$2714(\Lambda\Xi\pi\pi)$ \\ \hline\hline
$-6,0,0$&sc&$3334$&$-40$&$-11$&$0.1$&$1.1$&$
3282$&$-281$&$-62$&$0.4$&$3350$&$-30$&$5$&$0.1$&$1.0$&$3345(\Omega\Omega)$ \\ \cline{1-16}
$-6,0,1$&sc&$3355$&$-11$&$10$&$0.1$&$1.3$&$
3323$&$-118$&$-21$&$0.6$&$3348$&$0$&$4$&$0.0$&$3.1$& \\ \cline{1-16}
$-6,0,2$&sc&$3351$&$0$&$6$&$0.0$&$-$&$
3358$&$-11$&$13$&$1.2$&$3346$&$0$&$1$&$0.0$&$-$& \\ \cline{1-16}
$-6,0,3$&sc&$3346$&$0$&$1$&$0.0$&$-$&$
3346$&$0$&$1$&$-$&$3346$&$0$&$1$&$1.0$&$-$& \\ \hline
\end{tabular}
\newpage
Table VII: The same as Table III for $S=-4,-5$.

\begin{tabular}{c|c||c|c|c|c|c||c|c|c|c||c|c|c|c|c||c} \hline
\multicolumn{2}{c||}{Table VII}&
\multicolumn{5}{c||}{GBE(b)}& 
\multicolumn{4}{c||}{FN}&\multicolumn{5}{c||}{QDCSM} \\ \hline 
SIJ&&$E_{\alpha}$&$V_0$&$B_{\alpha}$&$\epsilon$&$R_s$&$E_{\alpha}$&$V_0$&$B_{\alpha}$&$R_s$&$E_{\alpha}$&$V_0$&$B_{\alpha}$&$\epsilon$&$R_s$&Threshold \\ \hline
$-4,0,0$&sc&$2654$&$-5$&$18$&$0.0$&$1.4$&$
2641$&$-39$&$5$&$1.0$&$2643$&$-24$&$7$&$0.1$&$1.2$&$2636(\Xi\Xi)$ \\ \cline{1-16}
&cc&$2654$&$-5$&$18$&$0.0$&$1.4$&$
2641$&$-40$&$5$&$1.0$&$2643$&$-24$&$7$&$0.1$&$1.2$& \\ \hline
$-4,0,1$&sc&$2806$&$-4$&$18$&$0.0$&$1.4$&$
2793$&$-38$&$5$&$1.0$&$2796$&$-18$&$8$&$0.1$&$1.3$&$2788(\Lambda\Omega)$ \\ \cline{1-16}
&cc&$2665$&$-26$&$29$&$0.1$&$0.9$&$
2641$&$-118$&$5$&$0.6$&$3016$&$-113$&$380$&$1.0$&$0.3$&$2636(\Xi\Xi)$ \\ 
\cline{1-16}
$-4,0,2$&sc&$2789$&$0$&$1$&$0.0$&$-$&$
2789$&$0$&$1$&$-$&$2789$&$0$&$1$&$0.0$&$-$& \\ \cline{1-16}
&cc&$2789$&$0$&$1$&$0.0$&$-$&$
2820$&$-4$&$32$&$1.1$&$2828$&$-50$&$40$&$1.0$&$0.7$& \\ \hline
$-4,0,3$&sc&$3073$&$-1$&$6$&$0.0$&$2.4$&$
3072$&$0$&$5$&$-$&$3071$&$-11$&$4$&$0.2$&$1.6$&$3066({\Xi}^{*}{\Xi}^{*})$ \\ \hline
$-4,1,0$&sc&$2653$&$-3$&$17$&$0.0$&$1.5$&$
2666$&$-93$&$30$&$0.6$&$2659$&$-3$&$23$&$0.1$&$1.3$&$2636(\Xi\Xi)$ \\ \cline{1-16}
&cc&$2655$&$-4$&$19$&$0.0$&$1.4$&$
2664$&$-95$&$28$&$0.6$&$2659$&$-4$&$23$&$0.1$&$1.3$& \\ \cline{1-16}
$-4,1,1$&sc&$2637$&$0$&$1$&$0.0$&$-$&$
2643$&$0$&$7$&$-$&$2637$&$0$&$1$&$0.0$&$-$& \\ \cline{1-16}
&cc&$2637$&$0$&$1$&$0.0$&$-$&$
2742$&$-17$&$106$&$0.6$&$2637$&$0$&$1$&$0.0$&$-$& \\ \hline
$-4,1,2$&sc&$2852$&$0$&$1$&$0.0$&$-$&$
2852$&$0$&$1$&$-$&$2852$&$0$&$1$&$0.0$&$-$&$2851(\Xi{\Xi}^{*})$ \\ \cline{1-16}
&cc&$2873$&$-13$&$21$&$0.1$&$1.1$&$
2893$&$-43$&$41$&$0.7$&$2862$&$-74$&$10$&$1.0$&$0.7$& \\ \hline
$-4,1,3$&sc&$3063$&$0$&$6$&$0.0$&$-$&$
3061$&$0$&$4$&$-$&$2999$&$-96$&$-58$&$0.8$&$1.0$&$3057({\Sigma}^{*}\Omega)$ \\ \cline{1-16}
&cc&$3068$&$-12$&$11$&$0.1$&$1.3$&$
3073$&$-3$&$16$&$1.4$&$3011$&$-84$&$-46$&$0.6$&$1.0$& \\ \hline\hline
$-5,1/2,0$&sc&$3207$&$0$&$1$&$0.0$&$-$&$
3100$&$-252$&$-106$&$0.5$&$3199$&$-37$&$-6$&$0.2$&$1.1$&$3205({\Xi}^{*}\Omega)$ \\ \cline{1-16}
&cc&$3213$&$0$&$7$&$0.0$&$-$&$
3100$&$-252$&$-106$&$0.5$&$3199$&$-37$&$-6$&$0.2$&$1.1$& \\ \hline
$-5,1/2,1$&sc&$3001$&$-7$&$16$&$0.1$&$1.3$&$
2977$&$-63$&$-14$&$0.9$&$3007$&$-7$&$17$&$0.1$&$1.3$&
$2990(\Xi\Omega)$ \\ \cline{1-16}
&cc&$3013$&$-10$&$22$&$0.1$&$1.1$&$
3013$&$-136$&$22$&$0.5$&$3125$&$-24$&$134$&$1.0$&$0.5$& \\ \cline{1-16}
$-5,1/2,2$&sc&$2991$&$0$&$1$&$0.0$&$-$&$
2991$&$0$&$1$&$-$&$2991$&$0$&$1$&$0.0$&$-$& \\ \cline{1-16}
&cc&$2998$&$0$&$7$&$0.0$&$-$&$
3003$&$-3$&$13$&$1.6$&$2995$&$0$&$4$&$0.0$&$-$& \\ \hline
$-5,1/2,3$&sc&$3207$&$0$&$1$&$0.0$&$-$&$
3207$&$0$&$1$&$-$&$3207$&$0$&$1$&$0.0$&$-$&$3205({\Xi}^{*}\Omega)$ \\ \cline{1-16}
&cc&$3213$&$0$&$7$&$0.0$&$-$&$
3212$&$0$&$6$&$-$&$3216$&$-4$&$11$&$0.2$&$1.6$& \\ \hline
\end{tabular}
\newpage
FIGURE CAPTIONS\\
Fig.1 Channel coupling effect for GBE(a) and (b) where the state is denoted 
by the symbols S,I,J for spin, isospin and angular momentum.
The single channel (sc) case for GBE(a) is shown as the solid curve;
the channel coupling case (cc) for GBE(a) is shown as the thick solid curve;
the single channel (sc) case for GBE(b) is shown as the dashed curve;
the channel coupling case (cc) for GBE(b)is shown as the thick dashed curve.\\
Fig.2 The effect of different smearing forms for GBE models.
The solid curve is for the GBE(a) model;
the dashed curve is for the GBE(b) model.\\
Fig.3 Additional channels as in Fig.2\\
Fig.4 States with similar pure repulsive effective interactions for the GBE, 
FN and QDCSM models.  The solid curve is for the QDCSM; the dashed curve is 
for the GBE model and the dotted curve is for the FN model.\\
Fig.5 The same as Fig.4 for states with similar weak attraction for all 
three models. \\
Fig.6 The same as Fig.4 for states with similar effective interactions 
for the QDCSM and FN models. \\
Fig.7 The same as Fig.4 for states with similar effective interactions 
for the QDCSM and GBE models. \\
Fig.8 The same as Fig.4 for states with similar effective interactions 
for the GBE and FN models. \\
Fig.9 States with effective interactions in the QDCSM which lie between 
those of the GBE and FN model. The curves for each model are as in Fig.4. \\
Fig.10 States with strong intermediate-range attraction in the QDCSM, for
which $|V_0|$ decreases with the increasing strangeness. The curves for 
each models are as in Fig.4. \\
Fig.11 States for which $|V_0|$ increases with the increasing strangeness 
for  the FN model. The curves for the models are as in Fig.1. \\
Fig.12 H particle with single channel and channel coupling in the QDCSM, GBE 
and FN models. The curves for the models are as in Fig.4, and the thick 
curves are for the channel coupling case. \\

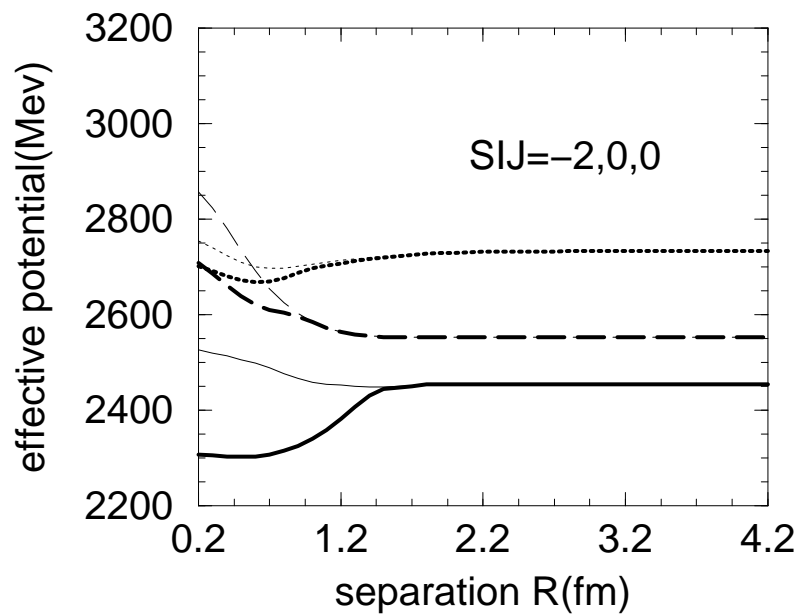
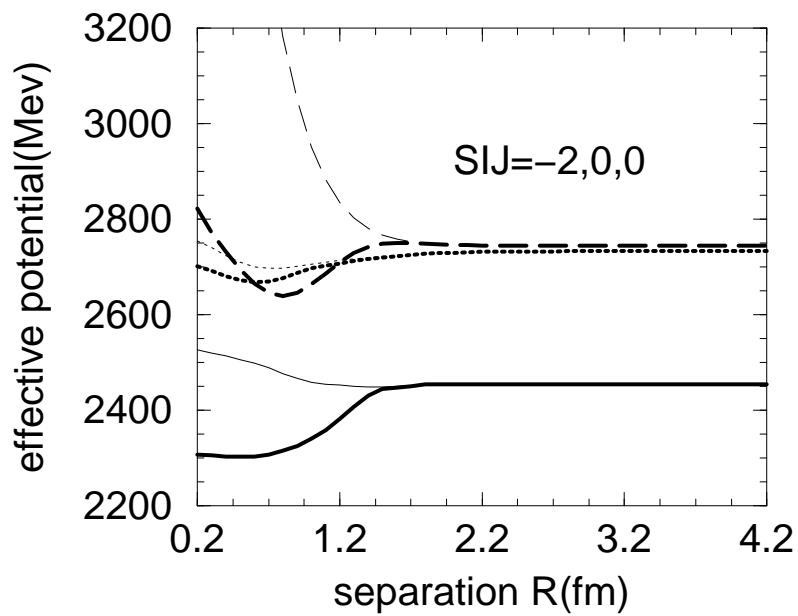
\end{document}